# Reservoir computing for sensing – an experimental approach

Dawid Przyczyna,[a,b] Sébastien Pecqueur,[c] Dominique Vuillaume[c] and Konrad Szaciłowski[a*]

[a]Academic Centre for Materials and Nanotechnology, AGH University of Science and Technology, Al. A. Mickiewicza 30, 30-059 Kraków Poland

[b]Faculty of Physics and Applied Computer Science, AGH University of Science and Technology, Al. A. Mickiewicza 30, 30-059 Kraków Poland

[c]Institut d'Électronique, Micro-électronique et de Nanotechnologie, CNRS, CS 60069, Av. Poincaré, 59652, Cedex, Villeneuve d'Ascq, France

*Corresponding author, e-mail: szacilow@agh.edu.pl

**Abstract**

The increasing popularity of machine learning solutions puts increasing restrictions on this field if it is to penetrate more aspects of life. In particular, energy efficiency and speed of operation is crucial, inter alia in portable medical devices. The Reservoir Computing (RC) paradigm poses as a solution to these issues through foundation of its operation – *the reservoir of states*. Adequate separation of input information translated into the internal state of the reservoir – whose connections do not need to be trained – allow to simplify the readout layer thus significantly accelerating the operation of the system. In this brief review article, the theoretical basis of RC was first described, followed by a description of its individual variants, their development and state-of-the-art applications in chemical sensing and metrology: detection of impedance changes and ion sensing. Presented results indicate applicability of reservoir computing for sensing and validating the SWEET algorithm experimentally.

# 1. Introduction

Problems of all civilizations arise and develop when the flow of one of the three fundamental components of systems is disturbed: matter, energy or information.[1] Insufficient resources or non-effective transportation of any of the above components can catastrophically affect prosperity of a regarded system. The rise and fall of complex civilizations as well as difficult situation in various regions of today's world can be explained on this basis. [1, 2] It is not surprising that there is intensive research carried out in each of these areas in the context of refining conventional solutions as well as discovering new, unconventional methods.

In information driven world, its effective acquisition and energy efficient processing is crucial. It seems, that silicon based computers are reaching their development potential, as miniaturisation reached its economical validity.[3] For this reason, many studies are being carried out on new hardware architectures for specific applications as well as on new, faster and more efficient software solutions.

Algorithms from the machine-learning paradigm [4], in particular neural network algorithms are gaining increasing interest in recent years. They are used in the topics such as quantum chemistry [5], pattern recognition [6], medical diagnosis [7], finance [8], social media [9], data mining [10] and even game-playing [11]. The structure of the classic neural network is depicted in Fig 1a.

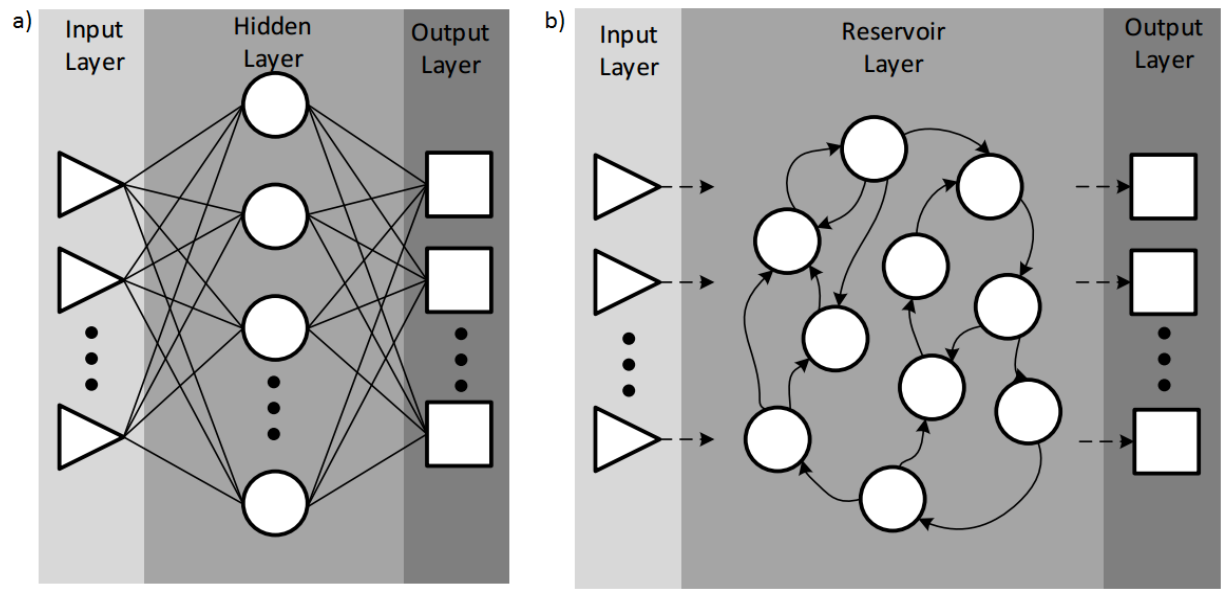

*Fig 1. Schematic structure of classical neural network divided into layers. The input layer, hidden layer and output layer are connected through connections which are optimized during the work of the algorithm. The share in transmission of data through connections (synapses) between nodes (neurons) is modifiable with its real-valued weight (a). Schematic framework of Reservoir Computing. Classically, Reservoir layer consist of randomly connected nodes with recurrent connections. Connections of inputs and nodes in reservoir are rigid, only connections to output layer are trained (b).*

When a neural network contains more than one hidden layer then it is called a deep neural network. Data flow through hidden layers is then a multistep process. Each layer can be trained on further refinements of features with an increasing complexity and abstraction of the output. Such network can be used in variety of topics, for example for searching new materials for photovoltaics [12], in autonomous vehicles [13], or for human-to-computer interface [14]. One of the breakthroughs in the field of machine learning was success of AlphaGo – computer program based on deep neural network and Monte Carlo Tree search, who managed to beat the world champion in go [11]. The strength of the algorithm was its recursive action, optimizing the steps taken in each move, simultaneously updating all possible paths as the game moved on. Achievements of such algorithms is undeniable, but it comes with a cost - to work fast, they require powerful supercomputing platform with huge energy inputs.

## 2. Reservoir Computing theory

In the area of recurrent neural networks (RNN), Reservoir Computing (RC) paradigm opens a new field for the development of software and hardware systems [15]. It addresses the problem of energy inefficient operation of RNN. RC operation principle is based on a non-linear dynamical system called *reservoir* (Fig 1b). Input layer feeds the data into reservoir layer where it is mapped onto high dimensional state space (Fig 2). Obtained versatility of states can be used for processing of non-linear dynamical systems with time-varying inputs and for linear separation of non-linear data through an increment of dimensionality [16]. Output signals consist of linear combination of internal states of reservoir at a given time, which depend on the system internal dynamics and the input data [15]. Therefore a reservoir computing system can be regarded, from the point of view of signal engineering, as a filter ($\mathfrak{F}$) [17] which maps the space of input time series $\Omega$ into the space of output time series $\Omega'$ (1):

$$\mathfrak{F}: \Omega \to \Omega' \tag{1}$$

If the configuration space of the reservoir is large enough and the input can move the system into various regions of this space, then the dynamic system can perform computation. In this context, the input layer can be understood as an interface translating the input signal into a form that can influence the internal dynamics of the reservoir, whereas the readout layer is an interface probing the internal state of the reservoir and extracting information which is stored there [17]. Huge advantage of RC is that connection weights are deliberately optimised only at the output layer, whereas reservoir layer is left to evolve freely according to its own dynamic properties and its interactions with the input data and/or the environment. In principle any dynamic system, which is responsive to some external stimulation can be used for reservoir computing. Good reservoirs (i.e. systems capable of information processing according to their internal dynamics and filter property) should provide separability (i.e. small difference in the input should result in significant change in the output and different inputs should result in different outputs), the echo state property (output layer reflects the internal dynamics of the system) and fading memory (the state of the reservoir depends on its history for a non-infinite period of time). This approach provides fast and energy efficient operation both for software and hardware solutions

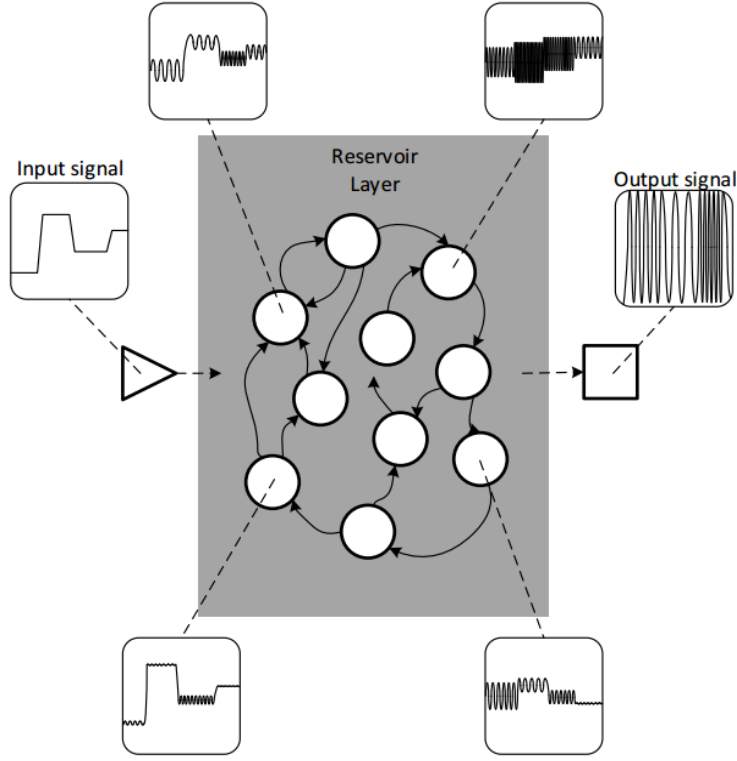

*Fig 2. Versatility of states of dynamical reservoir presented on the example of the frequency generation. Adapted from [18].*

Generally, by using a mathematical formalism, the operation of the reservoir can be described on the basis of a set of equations [19, 20]. The description of updating state of reservoir system is governed by discrete time equation:

$$x(n+1) = (1-a)x(n) + f\left(\mathbf{W}x(n) + \mathbf{W}^{in}u(n+1) + \mathbf{W}^{fb}y(n)\right), \quad (2)$$

where $n$ is discrete time, $x(n)$ is the N-dimensional reservoir state, $f$ is an activation function, $\mathbf{W}$ is the reservoir internal weight matrix, $\mathbf{W}^{in}$ is the input weight matrix, $u(n)$ is the K-dimensional input signal, $\mathbf{W}^{fb}$ is the output feedback matrix and **y(n)** is the L-dimensional output signal. $\mathbf{W}^{int}$, $\mathbf{W}^{in}$ and $\mathbf{W}^{fb}$ matrices determine connection among nodes. Commonly, the logistic sigmoid, hyperbolic tangent, rectifying linear unit functions are used as an activation function of neurons. The first part of the equation $(1-a)x(n)$ is added only if leaky-fire-and-integrate model of neurons is used for a different dynamic behaviour [15].

The output of the reservoir can be formulated as follows:

$$y(n+1) = \mathfrak{F}(\mathbf{W}) = f_{out}\left(\mathbf{W}^{out}\left(u(n+1), x(n+1), y(n)\right)\right), \quad (3)$$

where $y(n)$ is the readout output signals, $f_{out}$ is a linear continuous function, $\mathbf{W}^{out}$ is output weight matrix, and $\left(u(n+1), x(n+1), y(n)\right)$ is an integration of inputs, reservoir state and previous output vectors.

Readout of the reservoir is obtained by linear combination of reservoir states. Because state of each node will change upon inputs feed to the reservoir, the state matrix **S** is a combination of **K** input vectors and **N** nodes states matrices, giving $\mathbf{S} = \mathbf{N} \times \mathbf{K}$. As an operation principle of RC is based on multidimensional states of reservoir layer, $\mathbf{W}^{out}$ matrix weights are defined as combination of L-dimensions of N nodes states, $\mathbf{W}^{out} = \mathbf{L} \times \mathbf{N}$. The calculation of actual output $\mathbf{Y} = \mathbf{W}^{out} \times \mathbf{S}$ being as close to $\hat{\mathbf{Y}}$ desired "target" output as possible (in the least squared sense) is a linear problem; solution is based on Moore-Penrose pseudo inverse $\mathbf{S}^{+}$ of the **S** state matrix:

$$\mathbf{S}^{+} = \left(\mathbf{S}^{T}\mathbf{S}\right)^{-1}\mathbf{S}^{T} \tag{4}$$

$$\mathbf{W}^{out} = \left(\hat{\mathbf{Y}}\mathbf{S}^{+}\right)^{T} \tag{5}$$

On the other hand, common practice in obtaining a readout layer is usage of various regression function, e.g. Ridge, Least Mean Squares or Recursive Least Squares regressions [21]. The choice of proper algorithms depends on structure of the reservoir and the activation functions of neurons.

### 3. Computational Variants

Generally, RC comprises within its conceptual framework echo-state networks (ESN) and Liquid State Machine (LSM) algorithms that belong to the RNN group [15]. In a similar approach, computational structures can be realised on a delay-based system with feedback. [16] Additionally, the state weaving environment echo tracker (SWEET) algorithm has been recently proposed for sensing applications [22]. The main concepts are presented below along with the latest developments of algorithms along with state-of-the-art application strategies. Other approaches, although functional, due to niche applications (or other limitations) are not as widely used as ESN, LSM or delay-based algorithms and have not been reviewed.

*3.1. Echo state networks*

ESN algorithm first proposed in 2001 by Jaeger [19] is one of the two pioneering RC. Since then, a great amount of work has been done in regard of software and hardware implementations employing this computational variant. Recently, it has been used for various diagnostic purposes, e.g. for blood vessel segmentation in retinal images [23], dynamical detection of steady state visual evoked potentials [24], diagnosis of Parkinson's disease [25], determination of ADHD biomarker [26] or for fetal QRS detection [27]. Other applications regarding non-linear systems modelling are also explored [28, 29].

In this computational variant, neurons within reservoir layer are connected randomly and have recurrent loops in their connection pathways. Additionally, connection between neurons is sparse, commonly less than 20%. Activation function of neurons is generally a sigmoidal in character (e.g. tanh), but hybrid applications of Leaky integrator neuronal model had also been used [30]. States of dynamical reservoir are dependent on its history of inputs, so called *echo* of its states.

Essentially, the ESN reservoir is randomly created, but sometimes it can lead to poor performance with too much computational complexity on one side and too little computing power to reach the desired target on the other. To increase performance of ESN, Dingyuan *et al.* [28] proposed optimisation algorithm of the network structure by removing insignificant synapses. Modified ESN achieved better results in chaotic time series prediction than non-optimised ESN. In turn, Abdelbari *et al.* [31] proposed employment of evolutionary algorithms to optimise network design and achieved the intended results in modelling several different complex systems. Dolinský *et al.* [32] explored optimisation of a readout layer through locally regularized orthogonal forward regression. The results indicate that modified readout can improve robustness to noise and accuracy of ESN. On the other hand, Sun *et al.* [33] incorporated deep learning methodology in ESN modelling of time-series benchmarks. The obtained results indicate very good prediction accuracy (low normalized root mean-square error) with acceptable training time.

*3.2. Liquid State Machine*

The LSM proposed in 2002 by Maas [34] is other pioneering RC solution. Initially, LSM was developed on the basis of the neuroscience to explain the main computational properties of neuronal microcircuits [35]. The difference between ESN and LSMs are a different activation function of neurons in the reservoir: sigmoid activations are replaced with threshold functions (spiking or digital) – each neuron is modelled on the basis of "leaky fire-and-integrate" neuron model, so LSM is a type of spiking neural network [36]. Once the activation threshold is reached, neuron releases its energy to neighbouring neurons. This principle is similar to spatially distributed biological neurons [37]. The second difference is that LSM mainly employs biologically inspired structures in formation of reservoir layer. Recent advancements in regard to applications of LSN include: speech recognition [38], control of the movements of robots [39] and EEG seizure detection [40].

As improvements of the LSN algorithm, Schliebs *et al.* [41] showed that using dynamic activation function (high sensitivity for weak stimuli; low sensitivity for overstimulation) for spiking of neurons can improve the separation capabilities and efficiency of the network. On the other hand, Roy *et al.* [42] presented a new approach to the construction of the LSN structure along with an improved algorithm for the readout layer. Obtained results showed less errors and a better efficiency than other state-of-the art upgrades to LSN algorithm. Enhancement of computational capabilities of LSN has been shown by Li *et al.* in [43] by introduction of spike timing depended plasticity process along with intrinsic plasticity to the neurons in reservoir. Synchronous activation of neurons improved energy efficiency of LSN systems and improved computational performance due to self-organising property.

*3.3. Delay-based systems*

In order to meet the hardware challenges of too many components of the RC system, Appeltant *et al.* [16] and Pacquot *et al.* [44] have introduced an alternative approach in construction of the reservoir. Here, the reservoir layer consists of only one, nonlinear node operating in a delayed feedback mode, which fulfils the condition of minimizing complexity of necessary hardware for operation. The principle of operation of delay-based RC is based on *virtual neurons* (Fig 3). First, operations of sample and hold of an input signal are performed. Then, a so called masking operation is performed in order to obtain a higher versatility of states in the reservoir. The pre-processing by masking consists in modification of the input signal, which functions as diversification of the reservoir response, thus increasing its dimensionality in the state-space. The applied mask can take a form of digital binary values [45] or simple sinusoidal wave. It has been shown by Nakayama *et al.* that an analog mask with chaotic values is also efficient (Fig. 13) [46]. Multiplication of the input data with the masking function is the most commonly applied approach. Creation of virtual neurons involves dividing a given delay line into parts by time-multiplexing of obtained signal. Number of obtained virtual neurons is given by (6):

$$N = \frac{\tau}{\theta}, \qquad (6)$$

where $\tau$ is time duration of a delay line and $\theta$ is constant distance between each virtual neuron, i.e. the time interval in which output layer probes the state of the reservoir (Fig. 3). Based on the above, the structure of delay-based RC is classified as belonging to circular connection topology with discreet reservoir.

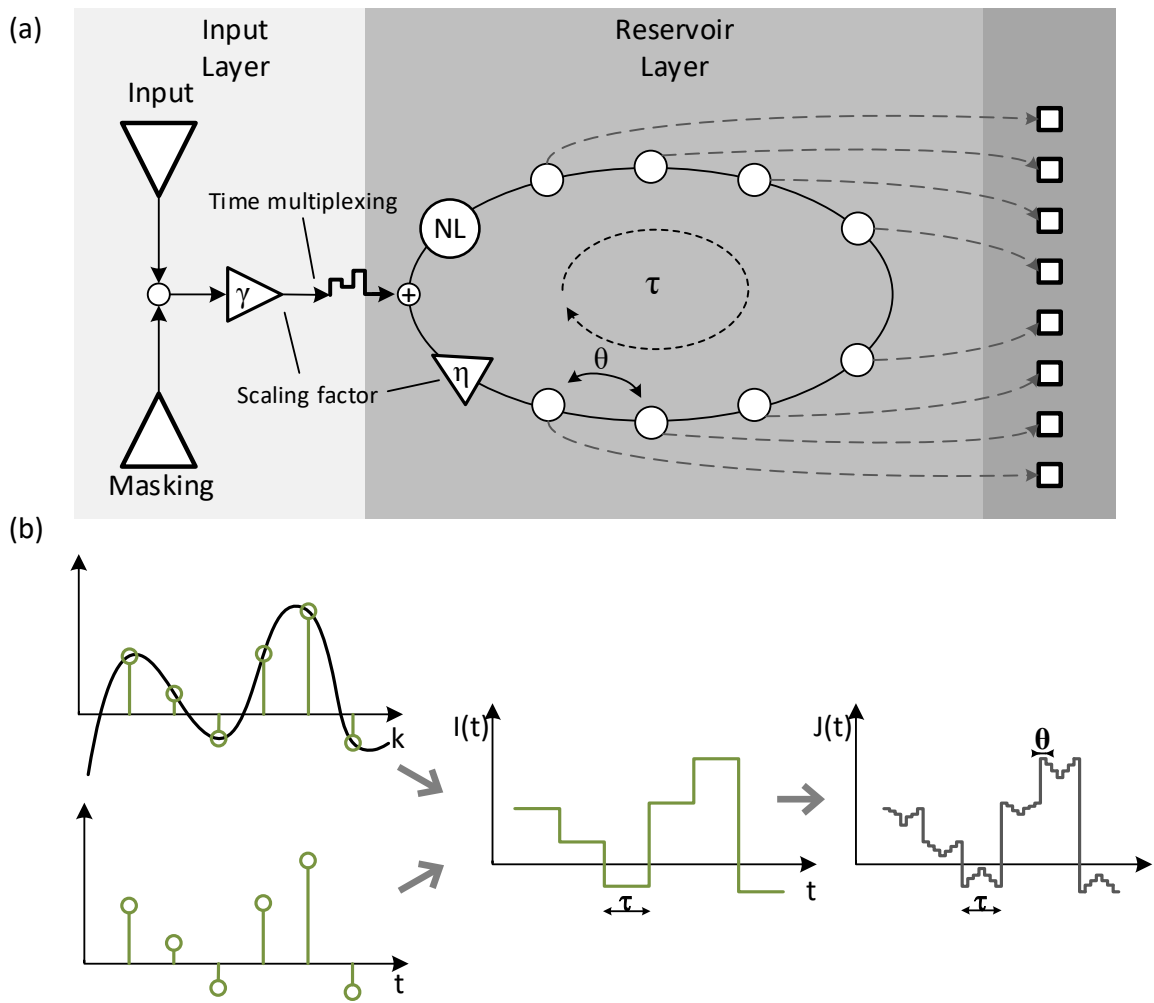

*Fig 3. A schematic structure of the delay-based reservoir computing device. τ is the length of the delay line, θ is the distance between neighbouring nodes. Masked input is scaled by scaling factor γ, as well as signal in the delay line, which is scaled by feedback strength η. Scaling operations are often required to ensure that the signals in the reservoir are not saturated nor supressed. Determination of scaling factors is optimised through numerical methods or*

*scanning through their different values to ensure proper system operation in regard to the given task or used non-linear node (NL). For the sake of simplicity, the output layer was constructed directly from the states of all virtual neurons. Adapted from [47] (a). Schematic visualisation of masking procedure. In the first place sampling of continuous (or discreet) input data is performed, which are then kept at a given level by the "hold" operation. After sample and hold operations, masking function is acting on the signal with specified time-multiplexed windows (θ). Those operations create virtual neurons (through time multiplexing) and ensure high versatility of their states (b). Adapted from [48].*

Due to limitation of operation speed of delay-based systems, they are mainly implemented in fast photonic systems. A state-of-the-art result in the topic of words classification task were shown by Larger *et al.* in [49]. Authors investigated two approaches to increase the functionality of the optical RC. Extended delay memory (EDM) procedure has been explored, which involved the scaling between the length of the input information and the duration of the delay line. This approach departed from the classical assumption that the input information must be equal to the length of the feedback loop. As a result, EDM enables temporal mixing of inputs in delayed line. Additionally, authors explored desynchronization of readout function (in relation to clock used to inject input information), proving numerically and experimentally the increase of functionality of reservoir by more than 1 order of magnitude. The investigated photonic RC enabled classification rate of stunning 1 million words per second.

On the other hand, Antonik *et al.* [50] examined the use of photonic RC in chaotic time series modelling and prediction. Authors presented first experimental hardware system capable of emulating Mackey-Glass series and Lorentz series with good agreement of Lyapunov exponents. However, the noise generated by system components (which are amplified in the feedback loop and weaken the accuracy of prediction after crossing specific threshold) is still a challenge to increase the functionality of the system.

**4. Implementation of reservoir computing for sensing**

Conventional chemosensing usually deals with noisy signals (electrical or optical) coming from sensors, i.e. sensing devices that have to respect figure of merits as an individual element (selectivity, signal-to-noise ratio, drift): In classical electrochemical sensors a change in electrode potential (in the case of potentiometric sensors, e.g. glass electrodes for pH measurements), electrode current (in the case of amperometric sensors, e.g. electrogalvanic oxygen sensors [51]). In the case of some chemosensors, the signal comes from light absorption or emission by specific sensor molecules [52, 53]. Some electrochemical sensors combine these two approaches. These sensors use electrodes modified with appropriate molecular species which contribute to the generation of analytical signal [48]. In the classic sensing approach, information about given environment is contained within variable response to stimuli of a given sensory platform, which is in contact with that environment. Comparison of the output data with the calibration curve gives us information about the content of the given environment. In an attempt to improve performance of a sensory platform, machine learning algorithms (including RC) can be used to enhance classification of its responses [54-56]. In this configuration, machine learning (or other advanced information processing routine) is used only as a post processing unit. The first step towards significant improvement of sensor performances (without engineering of the sensing material itself) was the so called heterodyne sensing [57-59]. In this sensing approach, frequency mixing in nanoscale materials induced by surface dipoles is recorded.

A completely novel approach was suggested recently by Zoran Konkoli [22]. The recently proposed SWEET algorithm adopts a reservoir which is weakly coupled with the environment under study. The difference can be highlighted in statements: '*reservoir computing **for** sensing*' in case of SWEET and '*reservoir computing **and** sensing*' for other approaches. This approach utilizes subtle changes in the internal dynamics of the reservoir due to interaction with the environment (i.e. the concentration or the flux of the analyte).

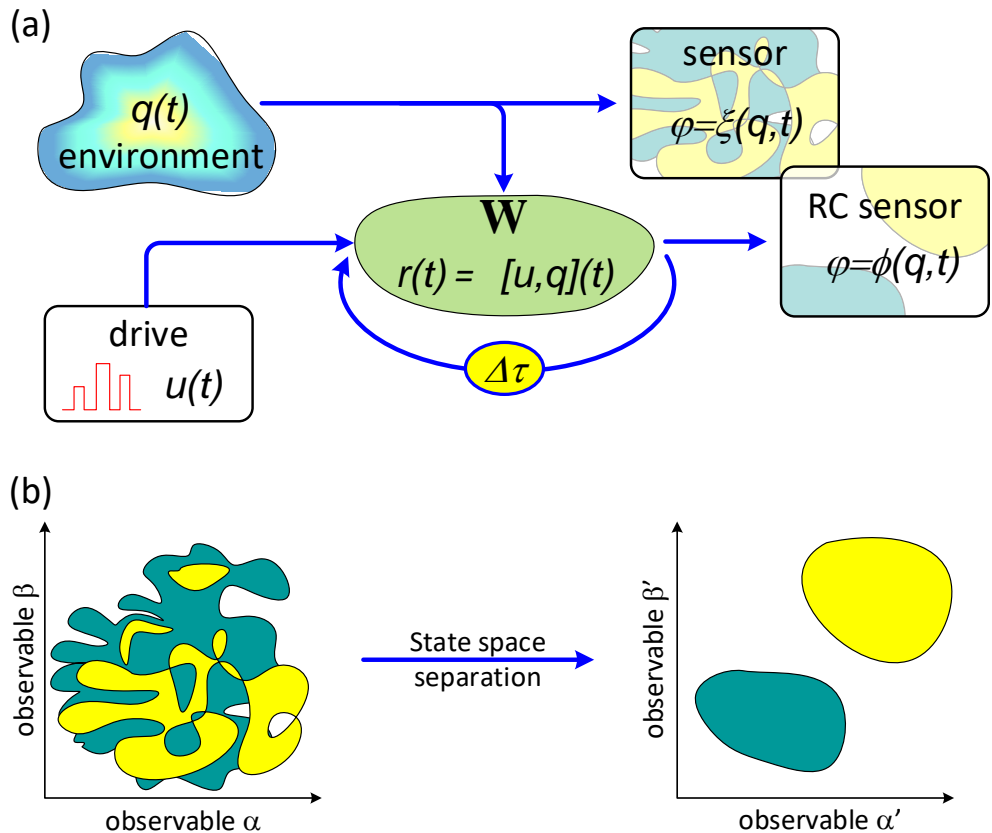

*Fig. 4. The overview of the SWEET sensing setup. The environment-sensitive reservoir is used to obtain information about the environment. This can be done on a direct pathway, which yields unprocessed information about the environment, ξ(q,t), or through indirect sensing by studying the response of the reservoir ϕ(q,t) to user-provided queries – the drive signal u(t). The delayed feedback Δτ is added to increase the complexity of the configuration space and achieve better reservoir computing properties (a). Symbolic representation of a complex sensing problem. A sensor has to detect the state of the system (yellow or green) on the basis of measurement of the observables α and β. On the left, the case of a low phase space separation in which the sensing problem is hard and prone to errors. Application of reservoir computing-based sensor with an appropriate drive signal performs domain separation of the system and yields much more reliable information. In this scenario, a relatively simple readout layer can be used to infer the state of the environment. Most of the computation is performed by the device, and not the readout layer. Adapted from Ref. [48].*

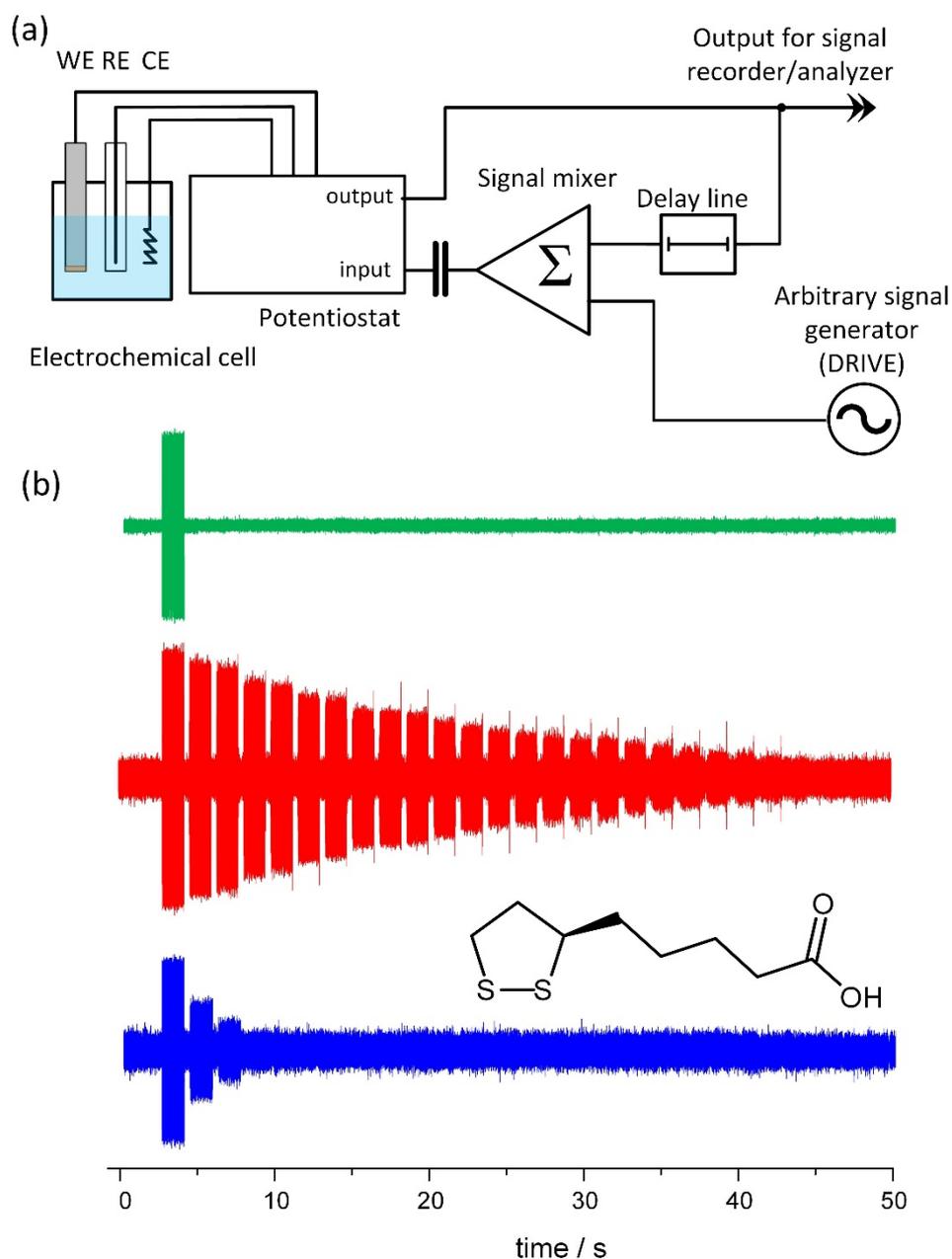

*Fig. 5. The block diagram of a reservoir sensor: WE stands for working electrode, RE for reference electrode and CE for counter electrode (a) and the response recorded with this device (b). The upper green curve presents the drive signal (2s, 100 Hz), the middle red one depicts the signal evolution in the absence of analyte (i.e. bare gold in potassium nitrate solution, pH = 7). The bottom blue curve presents the output of the device upon deposition of lipoic acid monolayer at the gold surface. Inset shows the molecular structure of the lipoic acid molecule. Potentiostat input controls the working electrode potential, the output is the analog current monitor.*

It can be argued that this is a new interpretation and extension of the RC paradigm, where intrinsic properties of the reservoir allow it to additionally act as a sensing platform and not merely as information processing unit. In the SWEET setup, the addition of a delayed feedback loop to the reservoir enhances its sensing potential due to the gradual build-up of even the smallest changes in its state, similarly to chaotic system. This means that the interaction of the reservoir with the environment can be very small, without compromising the quality of the detection. In principle, due to the high complexity of the reservoir-environmental layer, separation of internal states is achieved by adjusting input signal (drive) under different environment conditions. After finding proper drive that enables high separation of states in the phase space, relatively simple readout function suffices in probing reservoir internal states [22]. In contrast to classical sensing devices, the SWEET-based sensing devices can yield answers to any questions about the analyte concentrations – not only their actual values, but also their dynamic changes, or meeting a defined concentration threshold (Fig. 4).

The implementation of the SWEET concept was presented in a theoretical work in which the reservoir layer was constructed on the basis of one memristor (model was taken from literature) [60]. Basically, a memristor is a passive and non-linear circuit element, possessing switchable conductance states and non-volatile memory of its previous state [61]. A reservoir constructed this way was subjected to analysis and parameter optimisation. Two distinct drive signals were tested – one which was chosen arbitrarily based on scientific guess and the other optimised through *genetic algorithm*. Results show that both approaches allows for separation of different environmental conditions but the second approach is more robust towards potential scalability of memristor network constituting reservoir. It was discovered, that larger drive leads to faster operation at the cost of accuracy and should be optimised in regard to specific application. In addition, the authors introduced the "*Quality of sensing index v*", which can be used to determine the quality of sensing by examining the degree of separation of reservoir states in the phase space.

A first crude experimental verification of the reservoir computing for sensing approach was demonstrated to detect monolayers of lipoic acid on gold surfaces (Fig. 5). Lipoic acid reacts with gold surfaces forming stable monolayers. The impedance of these monolayers is higher than those of bare gold electrodes due to the higher charge-transfer resistance through the molecular layer. Application of a delayed feedback device (Fig. 5a) demonstrates the idea of reservoir computing for sensing. The increase of the impedance can be observed as a decrease of AC-signal amplitude (the first pulse of red vs blue tracks in Fig. 5b) or more convincingly, as an alteration of signal evolution in the feedback loop. Accumulation of small amplitude changes results in much faster signal decay in the system, so instead of amplitude changes, the analytical signal may be based on the time constant of the system, which means that the signal was transferred into a different domain.

Implementation of a similar type and architecture has also been developed on the basis of Organic Electrochemical Transistor (OECT) arrays covered with glycol-side-chain polythiophene. Instead than using networks of several OECTs as already reported [62], a unique OECT was used as single dynamical node operating in feedback loop employed for $K^+$ metal ions detection (Fig. 6). The channel of the OECT device was fabricated from a thin layer of PEDOT:PSS conducting polymer.

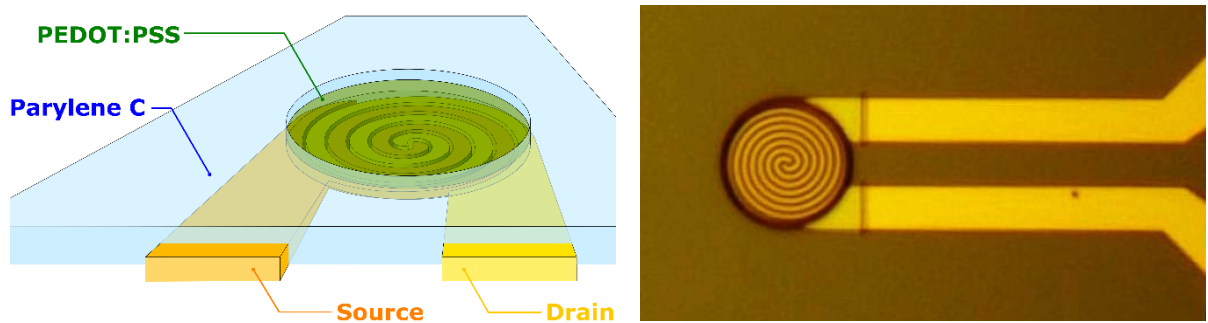

*Fig. 6. Design (left) and optical microscopy image (right) of PEDOT:PSS organic electrochemical transistor.*

The source-drain current was followed by a potentiostat in two electrode configuration, serving as an analog sourcemeter. Platinum wire placed in the electrolyte served as the gate of the transistor. The gate of the transistor was connected to the feedback loop containing also the drive generator (Fig. 7).

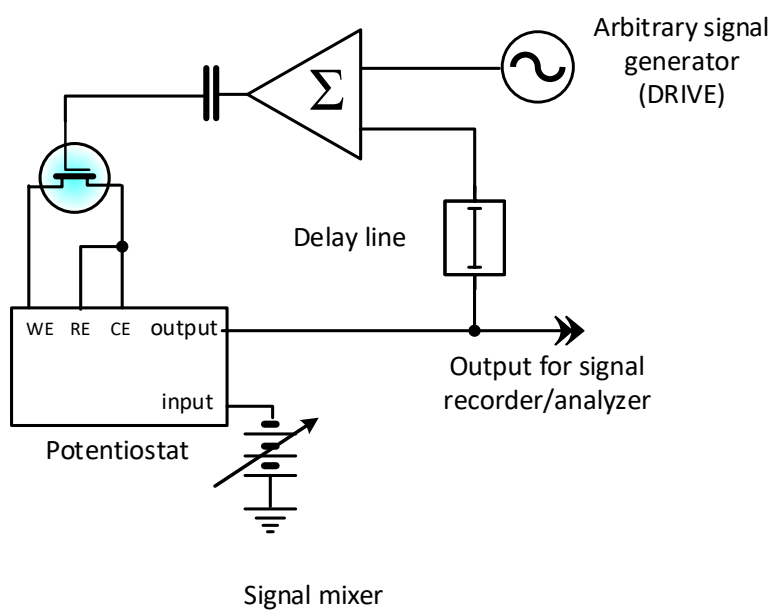

*Fig. 7. Connection diagram of a reservoir sensing system with OECT as an active element. Potentiostat input controls the potential source, the output is the analog current monitor.*

Current between source-drain (S-D) electrodes was forced by 100 mV DC potential. When the sinusoidal voltage (980 mV, 20 Hz, duration 1 s, interval between pulses 0.2 s) was applied to the gate electrode of OECT, ions present in the electrolyte were injected into to the channel, hence modulating the observed S-D current. Resulting signal was then delayed and applied to the gate electrode through the feedback loop with a capacitive coupling until the amplitude was insufficient to induce any response of the S-D electrodes (Fig. 8 a). Signal analysis was performed with Fourier transform to overcome noise present in the system. With increasing ion concentration, the obtained currents from the S-D electrodes were increased, which ultimately influenced the number of received signal packs. By plotting amplitudes of peaks from Fourier transforms versus repeated signal count, one can analyse time constant of fitted exponential curves, again, changing representation of the signal to a different state space (Fig 8 b). As a result, sensitivity of the system has been extended to higher concentrations, as it can be seen on the Fig 8 c. This promising preliminary results (Fig. 8 d) can be a starting point for more advanced studies of reservoir sensing with electrodes covered with specific molecular receptors or materials offering more complex dynamic behaviour (e.g. conducting polymers or semiconducting nanoparticles).

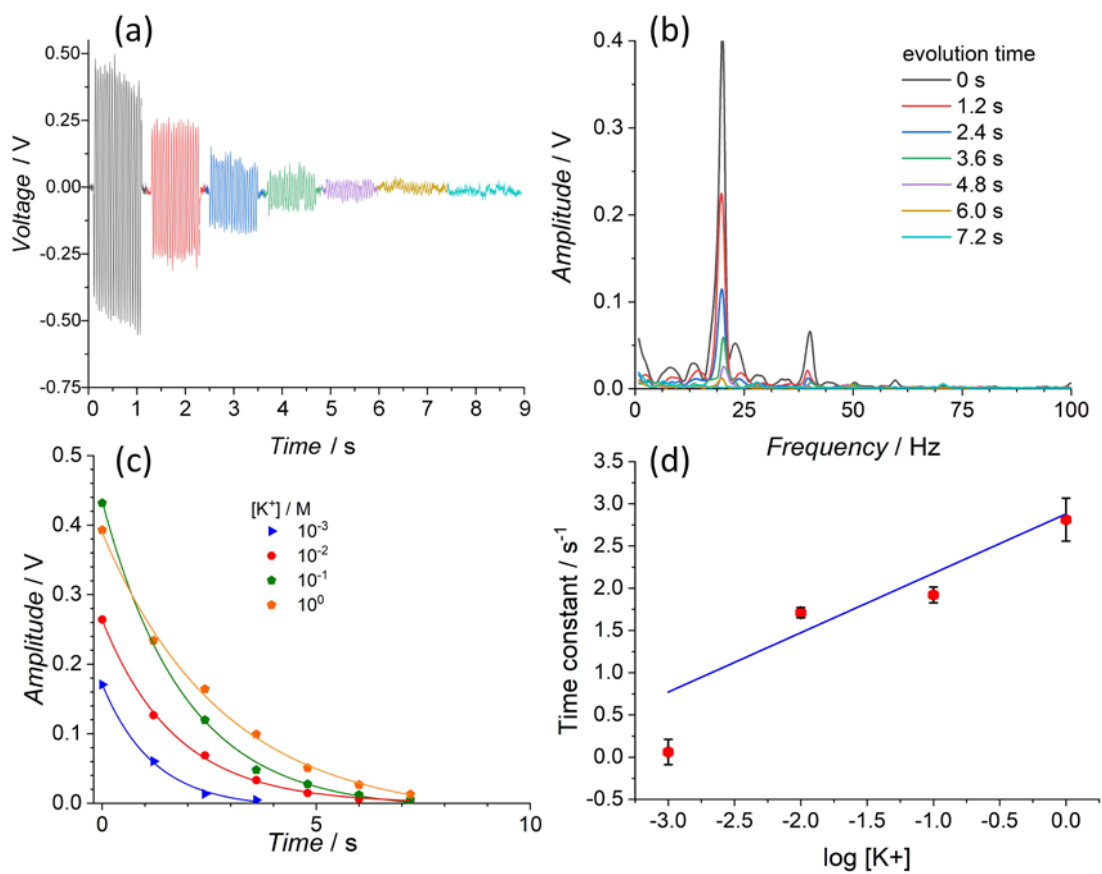

*Fig 8. Exemplary response of the studied reservoir computing sensing system with feedback loop at high $K^+$ ion concentration (a). Fourier transform spectra of each and every repeated signal package as a function of evolution time (b). Signal amplitudes of the first component (20 Hz) as a function of evolution time and potassium ion concentration (c), which enabled calibration of the time constants as a function of potassium concentration (d).*

The OECT-based reservoir sensing system again demonstrates the utility of the SWEET approach towards chemical sensing. The information on the concentration of the analyte was transduced by the ion sensitive polymer and subsequently transformed from a simple voltage output signal into an evolution of a dynamic system. Small nonlinearities in the circuitry resulted in a significant harmonic distortion of the drive and generation of higher harmonics. This in turns allows facile filtering of the drive signal and may contribute to increased accuracy of signal analysis. Furthermore, concomitant analysis of the evolution in time and frequency domains may be used to extract more information from electrochemical measurements.

## 5. Conclusions

Reservoir computing is a modern unconventional approach towards various problems associated with data classification, pattern recognition, and other applications that otherwise require high computational power. For numerous applications, it can be implemented in hardware. It is especially tempting to implement reservoir computing in various materials, e.g. assemblies of crystallites or nanoparticles, or thin molecular layers deposited at conducting substrates. Despite simplicity of the experimental approach, the computational performance may be useful for numerous practical applications, e.g. in chemical and biochemical sensing, environmental monitoring and other areas that may not necessarily require precise information of concentration, but rather binary classifications on specific thresholds of concentration.

## 6. Experimental

Lipoic acid-modified gold electrodes were prepared as follows. Gold disc electrodes (1 mm diameter) were polished with 1 μm alumina suspension on soft cloth, etched in Piranha solution for one minute, washed with deionized water and dried in air. 50 μl of saturated solution of lipoic acid in ethanol was poured over the electrode surface and left until all solvent evaporated. Subsequently each electrode was immersed in 100 ml of ethanol for 3 hours.

The fabrication of the OECTs was based on the methods described elsewhere [63, 64]: The Pt source and drain electrodes have an interdigitated spiral geometry for a total active surface of 615 μm² and 1.2 μm channel length [62]. The PEDOT:PSS polymer was spin-coated and structured using a lithographically-patterned Parylene mask [64, 65]. A first 2 μm-thick Parylene C layer was deposited as a passivation layer to prevent direct interfacing of the gold track with the analyte (3-acryloxypropyl trimethoxysilane was used as a adhesion promoted on the $SiO_2$ substrate). A second 2°μm-thick Parylene C layer was used to pattern the semiconductor as a mask (Micro-90 was used as a surfactant to prevent the adhesion between both Parylene-C layers. The structuring of the cavities over the active area of the OECT was performed at once through both Parylene-C layers by e-beam lithography and reactive ion etching. The PEDOT:PSS formulation was subsequently spin-coated on the device after a short UV-$O_3$ cleaning of the opened cavities.

Custom-built analog potentiostat "Kacper" (Instytut Fotonowy, Poland) was used in all experiments as a sourcemeter, with Ag/AgCl reference electrode and platinum wire counter electrode, which also served as a gate in OECT experiments. 0.1 M $KNO_3$ was used as a supporting electrolyte. G-Major/2 signal processing device (TC Electronics, Denmark) was used as a delay line, whereas audio signal mixer ZED Sixty 14 FX (Allen&Heath, UK) was used as mixing amplifier. Signals were recorded on Crimson 3 audio USB interface (Sound Performance Lab, Germany). Arbitrary function generator TG2512A (Aim-TTI, UK) was used as a source of a drive signal.

## Acknowledgements

Authors thank dr. Ewelina Wlaźlak for her help during SWEET experiments and data analysis. Furthermore, authors thank professor Kapela Pilaka for his friendly support, spiritual atmosphere in our laboratory and numerous discussion over sushi. The authors wish to acknowledge a series of discussions with all the partners of the RECORD-IT consortium. This work has received funding from the European Union's Horizon 2020 research and innovation programme under grant agreement No. 664786 and National Science Centre (Poland) under grant agreement No. UMO-2015/18/A/ST4/00058. DP has been partly supported by the EU Project POWR.03.02.00- 00-I004/16.## 6. Literature